\documentclass[conference]{IEEEtran} \IEEEoverridecommandlockouts
\usepackage{cite} \usepackage{amsmath,amssymb,amsfonts}
\usepackage{fancyhdr}
\usepackage{algorithmic} \usepackage{graphicx} \usepackage{textcomp}
\usepackage{subcaption}
\usepackage{tikz}
\usepackage{listings}
\usepackage[T1]{fontenc}
\usepackage[utf8]{inputenc}
\usetikzlibrary{arrows.meta,positioning,fit,backgrounds,calc}
\usepackage[pdftex,dvipsnames]{xcolor} \usepackage{orcidlink}
\usepackage{cleveref} \usepackage{braket}
\usepackage{booktabs}
\usepackage{tikz}
\usepackage{tikzit}

\tikzstyle{lalign}=[right]
\tikzstyle{ralign}=[left]

\tikzstyle{machine ibm}=[-, fill={rgb,255: red,207; green,190; blue,236}]
\tikzstyle{fugaku}=[-, fill={rgb,255: red,199; green,233; blue,255}]
\tikzstyle{quantinuum}=[-, fill={rgb,255: red,192; green,255; blue,242}]
\tikzstyle{border}=[-,dotted, thin]
\tikzstyle{box}=[-,densely dashed, very thick, blue]
\tikzstyle{network}=[<->, thick]
\tikzstyle{flow}=[->]

\usepackage{fontawesome5}
\usepackage{xurl}
\DeclareMathOperator*{\argmax}{arg\,max}
\usepackage[colorinlistoftodos,prependcaption]{todonotes}

\def\BibTeX{{\rm B\kern-.05em{\sc i\kern-.025em b}\kern-.08em
T\kern-.1667em\lower.7ex\hbox{E}\kern-.125emX}}

\makeatletter
\newcommand{\linebreakand}{%
  \end{@IEEEauthorhalign}
  \hfill\mbox{}\par
  \mbox{}\hfill\begin{@IEEEauthorhalign}
}
\makeatother

\begin{document}

\title{Quantum-HPC Workflows Across Multiple Quantum Computing Platforms: Two Case Studies}

\author{ 
\IEEEauthorblockN{Philipp Seitz} \IEEEauthorblockA{\textit{Quantinuum Ltd.} \\ Cambridge, UK \\ \orcidlink{0000-0003-3856-4090}0000-0003-3856-4090}\and 
\IEEEauthorblockN{Miwako Tsuji} \IEEEauthorblockA{\textit{CCS, University of Tsukuba} \\ \textit{RIKEN R-CCS}\\ Japan \\ \orcidlink{0000-0003-4709-1969}0000-0003-4709-1969} \and
\IEEEauthorblockN{John Children} \IEEEauthorblockA{\textit{Quantinuum Ltd.} \\ Cambridge, UK \\ \orcidlink{0009-0000-8641-9168}0009-0000-8641-9168} \and
\IEEEauthorblockN{Yuta Kikuchi} \IEEEauthorblockA{\textit{Quantinuum Ltd.} \\ Tokyo, Japan \\ \orcidlink{0000-0002-1802-5260}0000-0002-1802-5260}\linebreakand 
\IEEEauthorblockN{Riku Masui} \IEEEauthorblockA{\textit{Quantinuum Ltd.} \\ Tokyo, Japan \\ \orcidlink{0000-0002-2481-0361}0000-0002-2481-0361}\and 
\IEEEauthorblockN{Juan Pedersen} \IEEEauthorblockA{\textit{Quantinuum Ltd.} \\ Tokyo, Japan \\ \orcidlink{0000-0002-2481-0361}0000-0002-2481-0361}\and
\IEEEauthorblockN{Kentaro Yamamoto} \IEEEauthorblockA{\textit{Quantinuum Ltd.} \\ Tokyo, Japan \\ \orcidlink{0000-0002-9994-1200}0000-0002-9994-1200}\and 
\IEEEauthorblockN{Ross Duncan} \IEEEauthorblockA{\textit{Quantinuum Ltd.} \\ Cambridge, UK \\ \orcidlink{0000-0001-6758-1573}0000-0001-6758-1573 }
}

\maketitle

\thispagestyle{fancy}
\lhead{}
\rhead{}
\chead{}
\lfoot{\footnotesize{
SC26, November 15-20, 2026, Chicago, Illinois, USA
\newline 979-8-3195-4789-7/26/\$31.00 \copyright 2026 IEEE}}
\rfoot{}
\cfoot{}
\renewcommand{\headrulewidth}{0pt}
\renewcommand{\footrulewidth}{0pt}


\begin{abstract}
        We describe the Quantum-HPC hybrid system comprising the supercomputer Fugaku, the QPUs Quantinuum Reimei and ibm\_kobe, and the Tierkreis hybrid workflow software.
        We present two demonstrations of the system at work.
        In the first, we compute the excited states of a biomolecular system using Fugaku and Reimei (simulator) to
        handle classical and quantum parts of the workflow respectively.
        In the second, Fugaku coordinates Reimei and ibm\_kobe working together in a combined VQE + QPE workflow to compute the ground state energy of a simple system.
\end{abstract}


\section{Introduction} \label{sec:introduction}


All practical implementations of quantum algorithms require some portion
of classical compute to perform tasks such as adapting the problem
statement to the quantum program, optimising the quantum program for the
hardware, or performing post-processing of results.

In recent years, due to the scarcity of quantum compute, implementations of
quantum algorithms have been proposed to take further advantage of classical
compute by the use of HPC systems to reduce the overall resource requirements for
solving particular problems \cite{farhi2014,tilly2022,kanno2026}. Typical
implementations of these quantum-HPC applications are bespoke experiments
built for specific hardware systems and their respective software environments.
Implementers find themselves having to consider facility-specific restrictions,
device characteristics, hardware limitations, and other unique circumstances,
e.g., networking topologies. As a result, an application workflow typically
includes a multitude of different programs, including schedulers and classical
methods, that are held together by glue code and scripts.

Recent successes in the space of hybrid applications for quantum chemistry
include quantum selected configuration interaction (QSCI)\cite{kanno2026}
and sample-based quantum diagonalisation (SQD)\cite{moreno2025}, which use
significant quantities of classical computation and therefore require the use
of HPC resources. QSCI, for example, makes use of the advantages of quantum
computing (sampling) and classical HPC (distributed diagonalisation) to obtain
accurate ground state energies. Similar techniques extend to other areas, for
example quantum mechanics/molecular mechanics (QM/MM) methods\cite{bickley2025},
or quantum machine learning\cite{koziellpipe2026}. As implementations of these
algorithms increase in size and complexity, and due to having to orchestrate
workloads across quantum and classical compute, application teams find
themselves having to spend more of their time maintaining the infrastructure
around their implementation rather than the algorithm itself.

Elsewhere, in the broader scientific community there is a trend towards
managing the complexity of scientific workflows through workflow management
systems~\cite{Suter_2026}. Workflow Management Systems (WMS) are widely adopted
in the cloud computing and machine learning spaces and as such many mature
general-purpose options such as Apache Airflow\cite{apache_airflow} or Prefect
\cite{prefect_software} are available. While general-purpose WMSs are generally
extensible through plugin software, they can lack specific capabilities that
are required for certain applications so some scientific communities will
make use of domain specific WMSs. For example, Snakemake\cite{koester2012} is
used for data analysis in bioinformatics, while Pegasus\cite{deelman2002} is
used for astronomy and physics. While there have been some efforts to adapt
Prefect\cite{prefect_software} and Covalent\cite{cunningham2023} to support
quantum workflows through \texttt{Qiskit} and \texttt{PennyLane} respectively,
these tools are not primarily designed to support quantum-HPC workflows.
In general, quantum-computing frameworks are still evolving and mostly targeting
a single device with exclusive access.
On a lower level, QDMI\cite{Wille2024} and QRMI\cite{Ohtani} provide a hardware interface.
The quantum computing community is yet to consolidate around a particular
tool or software stack for quantum-HPC integration.

In this work, we adapt the Tierkreis\cite{Sivarajah2022Tierkreis} quantum
workflow management system to the HPC environment.
It sits between quantum-computing frameworks (\texttt{Qiskit}, \texttt{PennyLane}, etc.) and the HPC environment
with generic or custom integrations to a quantum hardware interface.
We describe two hybrid
applications as they were implemented and executed on a combination of classical
and quantum computing hardware. The first experiment extends the classical
capabilities of Fugaku with the quantum capabilities of Reimei to compute the
excited states of a biomolecular system. Significant resources are necessary
for the classical segments as the workflow makes use of conventional HPC quantum
chemistry packages. Our implementation demonstrates a reusable workflow that
parallelises the quantum optimisation and submission. The second experiment
combines two quantum devices of different modalities, Reimei and ibm\_kobe, in
a single workflow. To our knowledge, this is the first end-to-end demonstration
of a multi-modality quantum-HPC workflow. In the experiment, we use ibm\_kobe to
warm-start a Quantum Phase Estimation (QPE) calculation on Reimei and improve on
a classically prepared initial state.

The scope of this work is the development and demonstration of a hybrid quantum-HPC workflow management system.
It demonstrates two end-to-end workflows integrating both classical HPC and quantum computing resources.
It is not a micro benchmark nor does it introduce a novel quantum algorithm; rather, it highlights future integration possibilities and practical considerations.
Specifically, we add support for HPC batch submission, quantum submission, and reusable worker interfaces.
The result is reusable workflows that can be easily adapted to different quantum and HPC backends.


\begin{figure}[b]
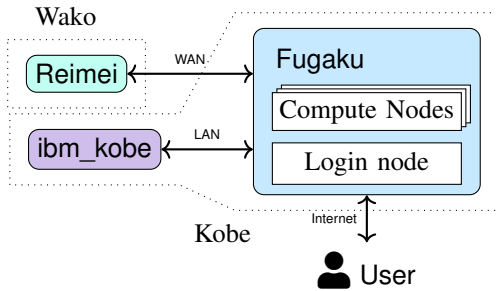

        \centering
        \ctikzfig{system}
        \caption{The Quantum-HPC System Overview.}
        \label{fig:system}
\end{figure}

\section{Quantum-HPC Hybrid System Architecture}\label{sec:system-architecture}

\Cref{fig:system} shows the hybrid system architecture.
Users can access Fugaku and the two quantum computers through one of its login nodes.
The HPC system can submit quantum jobs through low-latency connections.
More details about Reimei and ibm\_kobe are provided in \Cref{table:quantum-computers}.
The facility architecture informs the workflow design and implementation, which we will describe in \Cref{sec:tierkreis}.

\subsection{Fugaku}\label{sec:fugaku}
The supercomputer Fugaku is an
ultra-scale general-purpose manycore-based system with 158,976 nodes
\cite{sato2022fugaku}. The node specifications for Fugaku are summarised in
\Cref{table:fugaku-node}.
Fugaku employs a proprietary job scheduler from Fujitsu Technical Computing Suite (TCS) for batch job management\cite{sato2022fugaku}, which is optimised towards Fugaku's network topology.
\texttt{pjsub} is the job submission command for the Fujitsu job scheduler, corresponding to commands such as \texttt{qsub} and \texttt{sbatch} on other HPC systems.
\texttt{sqcsub} is a quantum job submission framework for submitting quantum
jobs through the direct connections between Fugaku and the two quantum
computers. A lightweight scheduler is deployed between Fugaku and each quantum
computer.
The SQC client library provides the corresponding APIs required to submit
quantum circuits from running HPC applications or workflow systems.

\subsection{Quantinuum Reimei}\label{sec:reimei}

Reimei is a 20-qubit trapped-ion quantum computer directly connected to the Fugaku supercomputer.
The Quantinuum device is using the System Model H1 design\cite{Pino2021}. It is located at the Wako
campus, approximately 500 km away from Fugaku, and the two systems were
interconnected through a virtual LAN.
\texttt{qsubmit} is a \texttt{gRPC} service that provides low latency
access to Quantinuum quantum computers. It provides a simple interface for
submitting quantum circuits and retrieving results.
It can be accessed through \texttt{sqcsub} on Fugaku.
Jobs submitted from Fugaku have the highest priority; if multiple HPC jobs are submitted simultaneously, a first-in-first-out
scheduling policy is applied. Beyond the direct connection, Reimei can also be
accessed through the cloud-based Quantinuum Nexus API.

\subsection{ibm\_kobe}\label{sec:ibm-kobe}
ibm\_kobe, a 156-qubit superconducting quantum computer, is located in the same building as Fugaku and directly connected to it through a low-latency network.
It is based on the Heron R2 architecture\cite{ibm_kobe}.
ibm\_kobe is available as a \texttt{Qiskit} backend.
Similar to Reimei, it can be accessed through direct submission integrated in \texttt{sqcsub} or through the cloud-based IBM Quantum API.

\begin{table}[]
        \caption{Quantum Resources in JHPC-Quantum}
        \centering
        \begin{tabular}{lll}\toprule
                                  & ibm\_kobe       & Reimei          \\\midrule
                QPU               & Heron R2        & System Model H1 \\
                Vendor            & IBM             & Quantinuum      \\
                Qubit Technology  & Superconducting & Trapped Ion     \\
                Qubits            & 156             & 20              \\
                Direct Access API & REST API        & gRPC            \\
                Site              & Kobe, Japan     & Wako, Japan     \\\bottomrule
        \end{tabular}
        \label{table:quantum-computers}
\end{table}

\begin{table}[]
        \caption{Specification of a single Fugaku compute node}
        \centering
        \begin{tabular}{ll}\toprule
                \multicolumn{2}{l}{\textit{Processor}}                     \\\midrule
                CPU         & A64FX                                        \\
                ISA         & Armv8.2-A with SVE                           \\
                Cores       & 4 groups of 12 (48 total)                    \\
                Frequency   & 1.8/2.0/2.2\,GHz                             \\
                SIMD Width  & 512\,bit                                     \\
                L1I/D Cache & 64\,KiB / 64\,KiB per core                   \\
                L2 Cache    & 8\,MiB per core group                        \\\midrule
                \multicolumn{2}{l}{\textit{Memory}}                        \\\midrule
                Type        & HBM2                                         \\
                Capacity    & 32\,GiB (8\,GiB per group)                   \\
                Bandwidth   & 1024\,GB/s                                   \\\midrule
                \multicolumn{2}{l}{\textit{Interconnect}}                  \\\midrule
                Network     & Tofu Interconnect D                          \\
                Bandwidth   & 40.8\,GB/s (28\,Gbps $\times$ 2 $\times$ 10) \\\bottomrule
        \end{tabular}
        \label{table:fugaku-node}
\end{table}

\section{Tierkreis Extensions for Quantum-HPC Systems}\label{sec:tierkreis}

Tierkreis\cite{Sivarajah2022Tierkreis} is a workflow management system designed
for the domain of quantum computing
\footnote{The original system was introduced for the use in cloud environments.
        Since then Tierkreis has been updated to support hybrid environments.}.
It assumes a workflow-centric pattern
~\cite{shehata2026quantumhpcsoftwarestacksopenqse} for hybrid applications,
where computations are broken down into independent and loosely-coupled
\emph{tasks}.
Workflows (programs) are implemented as directed acyclic graphs (DAGs) with
additional control flow and higher-order structures. As a result, independent
sections of the graph can be resolved in parallel. The Tierkreis controller can
leverage this information for orchestration and dispatch tasks to suitable
backends. In the quantum-HPC context, these backends can be classical, quantum, or
other specialised resources such as GPUs or FPGAs.
A task can be anything, from a simple integer addition to a large-scale HPC
application (e.g. an MPI-based program). Tasks may have any number of inputs
and outputs, which are managed by the Tierkreis runtime, without the need for
explicit handling by the user. Data moves through the workflow using batches
of input and output files at task boundaries. As the state of the workflow
(including inputs and intermediate values) is entirely persisted to disk in
this way, the workflow can be paused, resumed or restarted without loss of data
between tasks.
Tasks are provided by independent libraries called \emph{workers}. They consist
of task definitions and an implementation. A task definition is a typed
interface that describes the inputs and outputs of the task. The typed
interface is used to ensure correct usage of the task. Potential errors can
therefore be detected at workflow construction time, rather than at runtime.
In the HPC context, hardware resources can be classical, quantum, or other devices such as GPUs or FPGAs
which in Tierkreis are managed through \emph{executors}.
Worker tasks are associated with an executor, which handles the execution lifecycle.
Execution here can mean dispatching a local process, submitting a job to a scheduler,
or sending a quantum circuit to a quantum device. The executor is responsible
for providing the inputs to the task, kicking off the execution, and collecting the outputs.
Intermediate values (e.g. worker outputs) are stored and checkpointed in locations managed by Tierkreis.
The primary storage implementation in Tierkreis makes use of the shared filesystem in HPC context, but the storage interface
used by Tierkreis is abstract and other implementations are possible.
Additional information, such as metadata, error and execution logs, and task status are also stored along with the intermediate values in the same storage implementation to ease observability and debugging.
Workflows can be paused and resumed by suspending the Tierkreis runtime.
Workflows that encounter errors can be resumed from the last checkpoint after user intervention.
The Tierkreis runtime is the central component of the system.
It manages the global state of the workflow run, consisting of all node states and their dependencies.
It dispatches tasks through executors to the appropriate resources once all dependencies are satisfied.

\begin{lstlisting}[
  caption={Example of a Tierkreis workflow definition.},
  label={lst:tierkreis-example}
]
class InParams(NamedTuple):
    a: TKR[float]
    b: TKR[float]
    c: TKR[float]

g = Graph(InParams, TKR[float])
x = g.task(add(g.inputs.a, g.inputs.b))
y = g.task(add(x, g.inputs.c))
workflow = g.finish_with_outputs(y)
\end{lstlisting}

\begin{lstlisting}[
  caption={Example of a Tierkreis task definition.},
  label={lst:tierkreis-task-example}
]
@worker.task() # marks function as worker
def add(x: float, y: float) -> float:
    return x+y
\end{lstlisting}

The example workflow in Listing~\ref{lst:tierkreis-example} takes three floating-point inputs, $a$, $b$, and $c$, and constructs a graph consisting of two instances of the \texttt{add} task. The first task computes $x=a+b$, and its output is passed to the second task, which computes $y=x+c$. Thus, the workflow definition specifies the data dependencies between tasks and the final output, but does not specify how the tasks are executed.

The \texttt{add} task itself is defined independently in Listing~\ref{lst:tierkreis-task-example}. The \texttt{@worker.task()} decorator registers the function as a Tierkreis task, while its type annotations define the task interface: two floating-point inputs and one floating-point output. 

\subsection{Adapting Tierkreis for HPC Platforms}

Due to the unique characteristics of the system architecture, we adapt
Tierkreis to the HPC environment. We develop custom workers to handle the
software stack which we will describe in the following.
Tierkreis is designed as an abstraction layer, independent of the underlying
architecture. Therefore, integration with existing HPC platforms is
straightforward and mimics the process for local development environments. The
main difference is the choice of backends, which are responsible for executing
the tasks of the workflow.

We implement \texttt{pjsub} as an executor. In a workflow, users annotate HPC
tasks with resource descriptions (number of nodes, cpus, memory, etc.) and the
executors will handle submission and result retrieval. Quantum resources are
treated similarly; for real-time environments, they are part of the
resource definition exposed in the scheduler, for example, through the
QRMI\cite{bacher2025quantum}. Loosely coupled environments are managed by
\texttt{sqcsub} through a shell executor.

The shell executor is the generic interface for executing arbitrary tasks. In
the case of \texttt{sqcsub}, we use it in conjunction with a custom worker. We
declare the submission command as a worker interface; thus, the command can be
declared in the workflow graph and executed by the shell executor. Hardware
credentials and access controls are the responsibility of the host environment.
In this case, the SQC client library performs token-based user authentication by communicating
with the scheduler.

\subsection{Workflow Design}

Beyond the functional adaptation of Tierkreis to the HPC environment, we also consider the workflow design.
A primary advantage of using workflow systems is the composition of loosely coupled tasks.
Application developers can encapsulate functionality into one shared worker.
For example, one worker can handle the classical part of a hybrid algorithm, while another worker handles the quantum part.
As a result, they can develop the functionality which then can be reused in other workflows.
To achieve this, Tierkreis separates task definition from implementation; environment-specific implementations are possible.
The workflow can then be constructed solely from the task definitions without any knowledge of the underlying implementation.
Deploying the workflow in an environment consists of ensuring a suitable implementation is available for each task and selecting the appropriate executor for each task.
A second benefit is the ability to compose workflows from existing (sub-)workflows.
A common use-case is the compilation and submission of quantum circuits for different quantum backends.

\section{Computing Biomolecular Excited States}~\label{sec:comp-biom-excit}

%
Biomolecular excited states are electronic states of biological molecules, such as proteins and DNA, that result from their interactions with light. 
Computing such excited states can be computationally expensive when strong electron correlation must be treated in large active spaces. 
%
These calculations are important for understanding photochemical processes in biological and molecular systems.

Here, we describe a hybrid workflow for computing molecular excited states based on the work of Yamamoto et al.~\cite{yamamoto2026}. 
We reproduce the original Tierkreis workflow with minor modifications and omit error correction in the present experiments.
Due to a scheduled hardware upgrade for reimei, we run the experiments on the functionally equivalent reimei-simulator\footnote{This illustrates a practical benefit of the workflow system: adapting the experiment to a different backend requires a small configuration change.}.
Therefore, the experimental results may differ slightly from those reported in~\cite{yamamoto2026}.
The intention is to show the feasibility of the hybrid workflow, reproducing the key results despite changes in the experimental setup.

\subsection{Application Overview}

The total energy of the system is calculated using the subtractive ONIOM\cite{Spellmeyer2006} scheme:
\begin{equation}
        E_\text{system} = E_\text{high}^\text{model} + E_\text{low}^\text{real} - E_\text{low}^\text{model}
\end{equation}
The chosen low-level method is complete active space configuration interaction
(CASCI), and the high-level method is time-evolved quantum selected
configuration interaction (TE-QSCI). QSCI\cite{kanno2026} combines the advantages of
quantum computing (sampling) and classical HPC (diagonalisation). In
combination, both methods form a complex workflow that requires the use of both
classical and quantum resources. On the software side, the workflow combines
several existing software packages, including \texttt{NTChem}\cite{nakajima2015ntchem},
\texttt{TKET}\cite{sivarajah2021} and custom components for error correction and QSCI.
In the following, this paper focuses on the computer science and workflow framework aspects; for computational science methods and results, please refer to Yamamoto et al.\cite{yamamoto2026}.

\subsection{Workflow}

\begin{figure*}[t]
  \centering
  \input{figures/bio-wkflow.tex}
  \label{fig:biomolecular_workflow}
\end{figure*}

Figure \ref{fig:biomolecular_workflow} illustrates the overall hybrid workflow for computing molecular excited states.
First, an HF+CASCI calculation is performed on the HPC system.
In the example, the active space of $(6e,6o)$ corresponds to six electrons in six spatial orbitals.
We use \texttt{NTChem}\cite{nakajima2015ntchem}, for the initial molecular electronic structure calculation on Fugaku.
Then, for each target eigenstate, the workflow generates quantum circuits that approximately prepare the target eigenstate and perform time evolution.
We expand the configuration space to an $(8e,8o)$ active space, after which the generated time-evolution circuits are executed on the quantum computer Reimei to sample configurations.
Finally, the sampled configurations obtained from different target states and evolution times are merged to construct a common configuration space.
This is followed by subspace diagonalisation to obtain the final state energies, which is performed on Fugaku due to the computational requirements.

\subsection{Preliminary Experiments}

Here, we demonstrate a workflow that combines HPC and quantum computing resources.
For the quantum computing component, we use three backends: a local quantum simulator invoked directly from the workflow on a single core of the workflow server, the IBM quantum computer ibm\_kobe, and the Quantinuum reimei-simulator. The latter two are directly connected to Fugaku via a network.
Support for direct remote job submission was introduced as part of the HPC extensions in this paper, allowing users to execute these different workflows with only minimal modifications to their application code.

In this demonstration, the error correction for Quantinuum QPUs used in the original paper \cite{yamamoto2026}  is omitted to enable a direct comparison among multiple QPUs.
For HPC tasks (\texttt{NTChem}), 64 nodes of Fugaku are used.
The experiments use 4 MPI processes x 12 OpenMP threads per node on Fugaku.
All circuits use 16 qubits.
We use 200 shots for each of Qiskit Aer and reimei-simulator, and 20,000 shots for ibm\_kobe.
The shot counts differ because the backends had different access costs and latency profiles and execution fidelities.

The workflow engine is executed on a workflow server on Fugaku. 
The workflow server, like the login nodes, supports job submission to the job scheduler 
and is equipped with two Intel Xeon Gold 6338 processors (2.0 GHz, 32 cores each) and 256 GB of memory (16 × 16 GB DDR4-3200 RDIMM).
A single core is allocated to the workflow and each task.
The compiler, main libraries and operating systems are summarized in Table \ref{table:compilers-and-libraries}.


\begin{center}
\begin{table}[t]
\begin{center}
\caption{Compilers, libraries, and OSs}
\label{table:compilers-and-libraries}
\begin{tabular}{rr}\toprule
  Name   &  Version \\\midrule
  Compiler for \texttt{NTChem} & Fujitsu TCSDS 1.2.43 \\\midrule
  Python & 3.13.13 \\
  \texttt{tierkreis}& 2.0.12\\
  \texttt{pytket} & 2.18.0  \\
  \texttt{qiskit} & 2.4.1  \\\midrule
  OS for workflow Server&  RHEL 8.10\\
  OS for HPC nodes & RHEL 8.10 \\\bottomrule
\end{tabular}
\end{center}
\end{table}
    
\end{center}

\begin{table}[t]
        \caption{Workflow Execution Time (evolution time $= 7.5$) for the workflow in \Cref{fig:biomolecular_workflow} in seconds. Circuit Prep covers Freeze to compile.}
        \label{table:bm-exec-time}
        \centering
        \begin{tabular}{lrrrr}\toprule
                                        & \multicolumn{3}{c}{Fugaku}                               \\\cmidrule(lr){2-4}
                                        & Qiskit Aer        & reimei-simulator & ibm\_kobe  \\\midrule
                Total Exec. Time        & 3240              & 4066         &  3479                  \\
                \texttt{NTChem}         & 2653             & 2651       &   2689                    \\
                \texttt{NTChem} (Queue) & 29              & 53         &    98                \\
                Circuit Exec            & 0.5726          & 240.8      &     21.9            \\
                Circuit Prep            & 141.8          & 139.0      &     143.3       \\\bottomrule
        \end{tabular}
\end{table}

Table \ref{table:bm-exec-time} summarises the execution time of the entire workflow and its individual tasks.
The task execution time is measured from the start to the end of each task and therefore includes overheads such as preprocessing, postprocessing, and time spent waiting in the job queue. For reference, the queue waiting time is also shown for \texttt{NTChem}. In this experiment, the queue waiting time for execution is relatively short because a relatively small number of nodes are used. However, it should be noted that this waiting time depends on factors such as the workload and congestion of the supercomputer.
In the table, \texttt{NTChem} denotes the elapsed time from job submission to job completion, while \texttt{NTChem} (Queue) represents the waiting time in the queue.
The results for ``Circuit Exec'' and ``Circuit Prep'' are averaged over three tasks. ``Circuit Exec'' includes the entire elapsed time from job submission from Fugaku to the receipt of the results, including any queue waiting time and communication overhead.
A single workflow enabled the execution of \texttt{NTChem} on the supercomputer Fugaku,
the generation and execution of multiple independent circuits, and the merging of their results. 
The workflow is dominated by the execution time of the \texttt{NTChem} task, which accounts for approximately 80\% of the total execution time.
The circuit preparation also requires a non-negligible amount of time, suggesting that future optimisation and parallelisation of this stage will be necessary.
Table \ref{table:bm-energies} shows the excitation energies computed by CASCI of $(6e,6o)$ and $(8e,8o)$ and the TE-QSCI hybrid workflow.
For reference, the results of CASCI have been taken from \cite{yamamoto2026}; TE-QSCI values are reproduced.
The main benefit of Tierkreis is the reduction of orchestration complexity and verbosity combared to maintaining bash scripts.
This is particularly apparent when workflows are rerun with different backends, e.g., switching from reimei to reimei-simulator changes a single line.

\begin{table}[t]
        \caption{Excitation Energies (eV)}
        \label{table:bm-energies}
        \centering
        \begin{tabular}{llrr}\toprule
                Method               & Active Space & $T_0$ & $S_1$ \\\midrule
                CASCI                & $(6e,6o)$    & 1.44  & 1.48  \\
                TE-QSCI (ibm\_kobe)  & $(8e,8o)$    & 1.29   & 1.26     \\
                TE-QSCI (reimei-simulator) & $(8e,8o)$    & 1.26  & 1.23  \\
                TE-QSCI (Aer)        & $(8e,8o)$    & 1.28  & 1.30  \\
                CASCI                & $(8e,8o)$    & 1.26  & 1.29  \\\bottomrule
        \end{tabular}
\end{table}

\section{Combining QPU Modalities}
\label{sec:combining-qpus}

In our second case study, we demonstrate the coordination of both
quantum processors, Quantinuum Reimei and ibm\_kobe, on a single
computation.  Each QPU is used for a task that plays to its strengths,
achieving a better result than either alone.
The focus is again on the orchestration of the hybrid workflow;
the devices are not physically connected. It is an demonstration
of a heterogenous, modality-aware workflow.

We compute the ground-state energy of a 16-site transverse-field Ising
model (TFIM) defined by the Hamiltonian
\begin{equation}
        H = -J\sum_{\langle i,j \rangle}Z_i Z_j - h \sum_{i}X_i
        \label{eq:tfim}
\end{equation}
on a square lattice with periodic boundary conditions. $X_i, Z_i$ are
the Pauli operators acting on the spin at site $i$ and the indices
$\langle i,j \rangle$ represent pairs of nearest neighbours.

Our preferred algorithm for this task is quantum phase estimation (QPE)
\cite{kitaev1995quantum,Cleve:1997dh,Abrams1999,kitaev2002classical,Nielsen2010};
however, this algorithm requires relatively deep circuits, near the
limit of what Reimei can successfully execute.  To increase the
overlap between the input state and the ground state, and hence the
chance of success, we ``warm-start'' the algorithm by using an input
state that is already close to the ground state.  We find this input
state using the variational quantum eigensolver (VQE) algorithm
\cite{Peruzzo:2014aa}.  Since VQE typically takes many rounds to
converge, and each round needs many shots, we run this part of the
computation on the comparatively fast ibm\_kobe.  A high-level view of
the total workflow is shown in Figure~\ref{fig:wholeworkflow}.

We use VQE because, when fully converged, it provides a
\emph{classical} description of the final state itself, via the ansatz
and its optimal parameters.  This allows the same quantum state to be
prepared on the other QPU for use in the QPE algorithm, without the
need for quantum communication.

\subsection{Experiment Design}

\subsubsection{VQE} 
\label{sec:vqe}

In the first phase of the experiment, we approximate the ground state
of the Hamiltonian \eqref{eq:tfim} using VQE.  We minimise the energy
expectation value
\begin{equation}
        E(\theta) = \langle\psi(\theta)|H|\psi(\theta)\rangle
\end{equation}
by varying the variational parameters $\theta$ of the
trial state $\ket{\psi(\theta)}$.  We selected a parameterised Hamiltonian variational
ansatz~\cite{wecker2015}. The
ansatz circuit for $\ket{\psi(\theta)}$ takes the form
\begin{equation}
  \prod_{d=1}^D(\prod_i R_{Z_i}(\lambda_d)R_{X_i}(\phi_d)\prod_{\langle i,j \rangle}R_{Z_iZ_j}(\theta_d))(R_Y(\alpha)\ket{0})^{\otimes n} \,
\end{equation}
where the uniform product state is classically
pre-optimised over a single parameter $\alpha$.  The variational
parameters ${\{\lambda_d,\phi_d,\theta_d\}}_{d=1}^D$ are shared across
all qubits for each layer, reflecting the translational and rotational
symmetry of the lattice.

The parameter optimisation uses the simultaneous perturbation
stochastic approximation (SPSA)~\cite{spall1992,spall1998}, which
evaluates the energy at two points on each iteration; since the
Hamiltonian contains $X$ and $ZZ$ terms, we need to measure in both
bases, for a total of four measurements.
We use the \texttt{InQuanto}\cite{inquanto} implementation of SPSA.


\subsubsection{QPE}
\label{sec:qpe}

The second phase of the experiment computes the ground state energy of
$H$ using Quantum Phase Estimation.  Given the unitary $U = e^{i\lambda H}$ with the
eigenstates ${\{\ket{\phi_j}\}}_{j=1}^{N}$ and the initial state
\begin{equation}
  \ket{\psi} = \sum_{j=1}^{N}a_j\ket{\phi_j}\,
\end{equation}
we estimate a subset of the corresponding eigenphases
${\{\phi_j\}}_{j=1}^{N}$. When $\ket{\psi}$ approximates the ground
state, $\phi_0/\lambda$ is most likely the ground state energy.  We
use VQE as described above to find a good candidate $\ket{\psi(\theta)}$.

We use the Information Theory QPE algorithm~\cite{svore2014}, since it
uses only a single ancillary qubit and a relatively shallow circuit.
This circuit consists of four components: preparation of the
input state $\ket{\psi}$ and then ancilla $\ket{+}$; application of $U^k$ on
the system register, conditioned on the ancilla qubit, where $U$ approximates
the time evolution operator $e^{i\lambda H}$; application of a phase-shift gate
$R_Z(\beta) = e^{-i\frac{\beta}{2}Z}$; and measurement of the ancilla qubit in
the $X$-basis.

We repeat the circuit $N_S$ times, sampling the parameters ${\{k_l\}}$
and ${\{\beta_l\}}$ uniformly from
$k\in{\{1,2,\ldots,k_\text{max}\}}$ and
$\beta \in {\{0,\frac{\pi}{2}\}} $, respectively, where $k_\text{max}$
controls the precision of QPE, and determines the maximum depth of the
circuit.

We can estimate $\phi_0$ by the phase $\tilde{\phi}$ that
maximises
\begin{equation}
  \tilde{\phi_0}=\argmax_{\tilde{\phi}}\prod_{l=1}^{N_S}\frac{1+d_{k_l}\cos{(k_l\tilde{\phi}+\beta_l - m_l\pi)}}{2}\,
\end{equation}
with the number of shots $N_S$ and the collected measurement outcomes
${\{m_l\}}_{l=1}^{N_S}$. The accuracy of this estimate increases with the
parameters $k_{\text{max}}$ and $N_S$, and the population ${|a_0|}^2$ of the
ground state $\ket{\phi_0}$ in the input state $\ket{\psi}$.

We optimise the circuit performance using \textit{circuit compression}
exploiting the  symmetry  $QUQ=U^\dagger$ in the
Trotterised evolution operator $U$, \textit{coherent noise
  suppression} by periodically inserting Pauli $X$ throughout the
circuit~\cite{viola1999}, and \textit{leakage detection} by appending
leakage gadgets~\cite{Stricker2020} on the system (16 qubits) and ancilla (1 qubit) register
and post-selecting on their measurement outcomes.  These leakage
detections use Reimei's remaining three qubits.

\subsection{Implementation Details}\label{ssec:implementation-details}

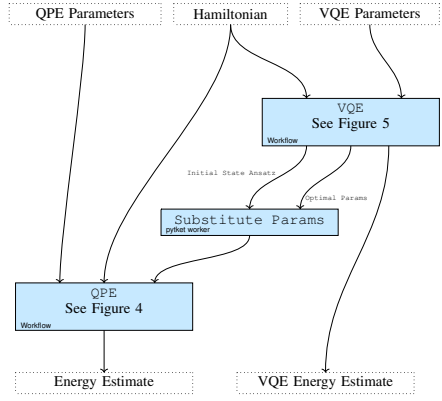
\begin{figure}[tbh]
  \centering
    \resizebox{0.33\textwidth}{!}{
      \begin{tikzpicture}
	\begin{pgfonlayer}{nodelayer}
		\node [style=none] (0) at (-5.75, 2) {};
		\node [style=none] (1) at (-8, 3) {};
		\node [style=none] (2) at (-3.5, 3) {};
		\node [style=none] (3) at (-3.5, 2) {};
		\node [style=none] (4) at (-8, 2) {};
		\node [style=none] (5) at (-5.75, 2.5) {Hamiltonian};
		\node [style=none] (6) at (-11.5, 2) {};
		\node [style=none] (7) at (-14.5, 3) {};
		\node [style=none] (8) at (-8.5, 3) {};
		\node [style=none] (9) at (-8.5, 2) {};
		\node [style=none] (10) at (-14.5, 2) {};
		\node [style=none] (11) at (-11.5, 2.5) {QPE Parameters};
		\node [style=none] (12) at (-0.25, 2) {};
		\node [style=none] (13) at (-3, 3) {};
		\node [style=none] (14) at (2.5, 3) {};
		\node [style=none] (15) at (2.5, 2) {};
		\node [style=none] (16) at (-3, 2) {};
		\node [style=none] (17) at (-0.25, 2.5) {VQE Parameters};
		\node [style=none] (18) at (-2, -14.5) {};
		\node [style=lalign] (23) at (-3, -6.25) {\tiny \tt Optimal Params};
		\node [style=none] (25) at (-5.5, -14.5) {};
		\node [style=none] (26) at (1.5, -14.5) {};
		\node [style=none] (27) at (1.5, -15.5) {};
		\node [style=none] (28) at (-5.5, -15.5) {};
		\node [style=none] (29) at (-2, -15) {VQE Energy Estimate};
		\node [style=ralign] (35) at (-3.75, -5) {\tiny \tt Initial State Ansatz};
		\node [style=none] (36) at (-1, -2) {{\tt VQE}};
		\node [style=none] (37) at (-4.5, -1.5) {};
		\node [style=none] (38) at (2.5, -1.5) {};
		\node [style=none] (39) at (2.5, -3.75) {};
		\node [style=none] (40) at (-4.5, -3.75) {};
		\node [style=lalign] (41) at (-4.5, -3.5) {{\tiny\sf Workflow}};
		\node [style=none] (42) at (-1, -3.75) {};
		\node [style=none] (43) at (-2.75, -1.5) {};
		\node [style=none] (44) at (-1, -1.5) {};
		\node [style=none] (45) at (1, -1.5) {};
		\node [style=none] (46) at (-1, -2.75) {{See Figure \ref{fig:vqe-wkflow}}};
		\node [style=none] (47) at (-5, -7.25) {{\tt Substitute Params}};
		\node [style=none] (48) at (-8.5, -6.75) {};
		\node [style=none] (49) at (-1.5, -6.75) {};
		\node [style=none] (50) at (-1.5, -8) {};
		\node [style=none] (51) at (-8.5, -8) {};
		\node [style=lalign] (52) at (-8.5, -7.75) {{\tiny\sf pytket worker}};
		\node [style=none] (53) at (-5, -8) {};
		\node [style=none] (54) at (-6.75, -6.75) {};
		\node [style=none] (55) at (-5, -6.75) {};
		\node [style=none] (56) at (-3, -6.75) {};
		\node [style=none] (57) at (-2.75, -3.75) {};
		\node [style=none] (58) at (0.5, -3.75) {};
		\node [style=none] (59) at (-10.75, -10.75) {{\tt QPE}};
		\node [style=none] (60) at (-14.25, -10.25) {};
		\node [style=none] (61) at (-7.25, -10.25) {};
		\node [style=none] (62) at (-7.25, -12.5) {};
		\node [style=none] (63) at (-14.25, -12.5) {};
		\node [style=lalign] (64) at (-14.25, -12.25) {{\tiny\sf Workflow}};
		\node [style=none] (65) at (-10.75, -12.5) {};
		\node [style=none] (66) at (-12.5, -10.25) {};
		\node [style=none] (67) at (-10.75, -10.25) {};
		\node [style=none] (68) at (-8.75, -10.25) {};
		\node [style=none] (69) at (-10.75, -11.5) {{See Figure \ref{fig:qpe-workflow}}};
		\node [style=none] (70) at (-12.5, -12.5) {};
		\node [style=none] (71) at (-9.25, -12.5) {};
		\node [style=none] (72) at (-10.75, -14.5) {};
		\node [style=none] (73) at (-14.25, -14.5) {};
		\node [style=none] (74) at (-7.25, -14.5) {};
		\node [style=none] (75) at (-7.25, -15.5) {};
		\node [style=none] (76) at (-14.25, -15.5) {};
		\node [style=none] (77) at (-10.75, -15) {Energy Estimate};
	\end{pgfonlayer}
	\begin{pgfonlayer}{edgelayer}
		\draw [style=border] (1.center) to (2.center);
		\draw [style=border] (2.center) to (3.center);
		\draw [style=border] (3.center) to (4.center);
		\draw [style=border] (1.center) to (4.center);
		\draw [style=border] (7.center) to (8.center);
		\draw [style=border] (8.center) to (9.center);
		\draw [style=border] (9.center) to (10.center);
		\draw [style=border] (7.center) to (10.center);
		\draw [style=border] (13.center) to (14.center);
		\draw [style=border] (14.center) to (15.center);
		\draw [style=border] (15.center) to (16.center);
		\draw [style=border] (13.center) to (16.center);
		\draw [style=border] (25.center) to (26.center);
		\draw [style=border] (26.center) to (27.center);
		\draw [style=border] (27.center) to (28.center);
		\draw [style=border] (25.center) to (28.center);
		\draw [style=fugaku] (40.center)
			 to (39.center)
			 to (38.center)
			 to (37.center)
			 to cycle;
		\draw [style=flow, in=90, out=-90] (0.center) to (43.center);
		\draw [style=flow, in=90, out=-90] (12.center) to (45.center);
		\draw [style=fugaku] (48.center)
			 to (51.center)
			 to (50.center)
			 to (49.center)
			 to cycle;
		\draw [style=flow, in=90, out=-90] (57.center) to (55.center);
		\draw [style=flow, in=90, out=-90] (42.center) to (56.center);
		\draw [style=flow, in=90, out=-90] (58.center) to (18.center);
		\draw [style=fugaku] (63.center)
			 to (62.center)
			 to (61.center)
			 to (60.center)
			 to cycle;
		\draw [style=flow, in=90, out=-90, looseness=0.50] (6.center) to (66.center);
		\draw [style=flow, in=90, out=-90, looseness=0.75] (0.center) to (67.center);
		\draw [style=flow, in=90, out=-90, looseness=0.75] (53.center) to (68.center);
		\draw [style=border] (73.center) to (74.center);
		\draw [style=border] (74.center) to (75.center);
		\draw [style=border] (75.center) to (76.center);
		\draw [style=border] (73.center) to (76.center);
		\draw [style=flow] (65.center) to (72.center);
	\end{pgfonlayer}
\end{tikzpicture}
    }
  \caption{Complete Workflow}
  \label{fig:wholeworkflow}
\end{figure}

The whole experiment is defined as a single workflow in Tierkreis,
and has two major parts: the VQE part and the QPE part.  Each of these
contains smaller subroutines to generate circuits, estimate the value
of Hamiltonians, and optimise classical parameters.  All of these
classical parts are also tasks within a single hierarchical workflow.


The general construction of the workflow is shown in
Figure~\ref{fig:wholeworkflow}.  The inputs ``QPE Parameters'',
``Hamiltonian'', and ``VQE Parameters'' are provided by the user when
starting the workflow.  The primary output is the result of the QPE
part.  However, since VQE is a subroutine in our complete workflow, we
can extract its outputs as an additional result, which we use as a
baseline for comparison.

\begin{figure}[thb]
  \centering
  \resizebox{0.5\textwidth}{!}{
    \input{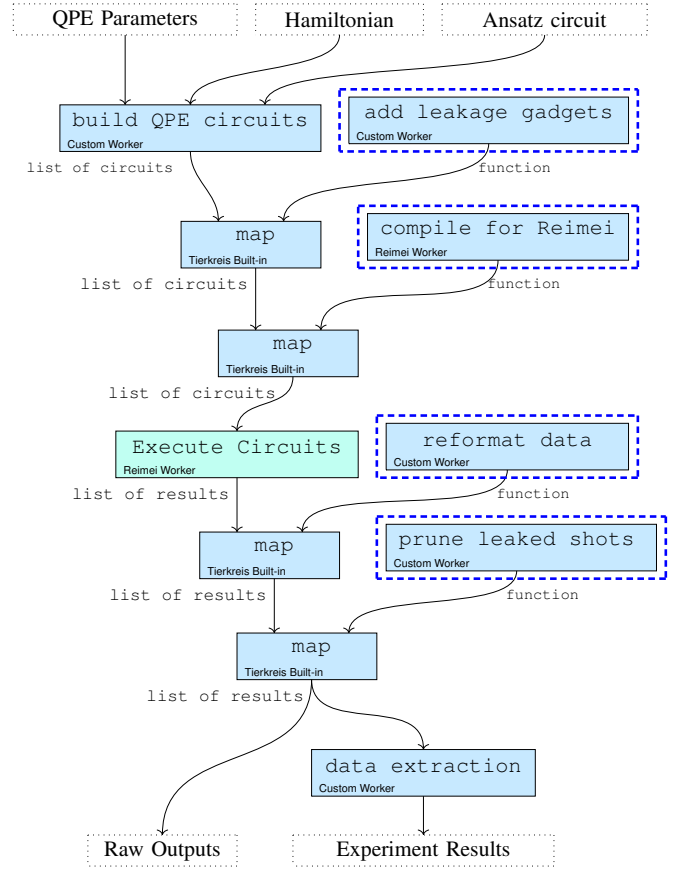}
  }
  \caption{
  QPE Workflow.  The dashed boxes indicate sub-workflows which are
  treated as data, to be used in the higher-order \texttt{map}
  operation.}\label{fig:qpe-workflow}
\end{figure}

We will describe the QPE part first, since it is simpler; it is shown in
Figure~\ref{fig:qpe-workflow}. The inputs to this workflow are the
Hamiltonian, the circuit used to prepare the initial state, and
various parameters for the algorithm.

The workflow has three phases: first, prepare the circuits for
execution using classical compute; second, execute on the Reimei QPU;
finally, process the results on the classical hardware to obtain
the experimental results.

The workflow appears largely sequential, but is notable for the use of
the higher-order \texttt{map} operation to uniformly apply an
operation to a list of operands.  This effectively transforms a list
of data into a list of tasks.  For example, mapping the \emph{compile
  for Reimei} operation over the list of circuits creates a
corresponding list of jobs; upon completion the results are
automatically reunited to produce the list of compiled circuits.
In principle, each of these tasks could be scheduled for concurrent
execution on different HPC nodes.  This would be overkill in our
experiment, so all these tasks were executed locally.

The ``raw results'' play no role in our analysis, but we retain
this data to validate that the experiment is working correctly.

\begin{figure}[thb]
  \centering
  \resizebox{0.5\textwidth}{!}{
    \input{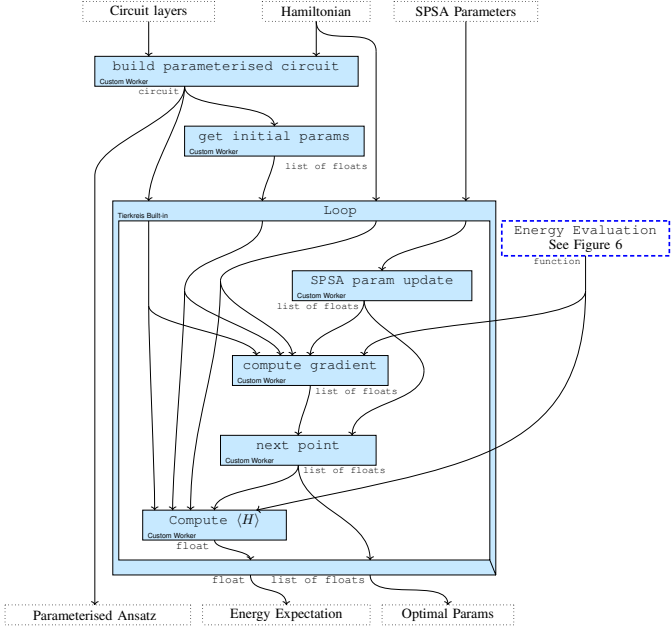}
  }
  \caption{VQE Workflow.  }\label{fig:vqe-wkflow}
\end{figure}

\begin{figure}[thb]
  \centering
  \resizebox{0.5\textwidth}{!}{
    \begin{tikzpicture}
	\begin{pgfonlayer}{nodelayer}
		\node [style=none] (0) at (-3.5, 6.5) {};
		\node [style=none] (1) at (-4.75, 5) {};
		\node [style=none] (2) at (-3.25, 7.5) {{\tt get QPU info}};
		\node [style=none] (3) at (-6.25, 8) {};
		\node [style=none] (4) at (-0.25, 8) {};
		\node [style=none] (5) at (-0.25, 6.5) {};
		\node [style=none] (6) at (-6.25, 6.5) {};
		\node [style=lalign] (7) at (-6.25, 6.75) {{\tiny\sf IBMQ Worker}};
		\node [style=lalign] (22) at (4.25, -1.25) {{\footnotesize\tt list of circuits}};
		\node [style=none] (26) at (3.75, -3.25) {{\tt map}};
		\node [style=none] (27) at (2, -2.75) {};
		\node [style=none] (28) at (5.5, -2.75) {};
		\node [style=none] (29) at (5.5, -4.25) {};
		\node [style=none] (30) at (2, -4.25) {};
		\node [style=lalign] (31) at (2, -4) {{\tiny\sf Tierkreis Built-in}};
		\node [style=none] (32) at (4.25, -1) {};
		\node [style=none] (33) at (4.25, -2.75) {};
		\node [style=ralign] (34) at (3.75, -4.5) {{\footnotesize\tt list of results}};
		\node [style=ralign] (56) at (-1.75, -1) {{\scriptsize\tt function}};
		\node [style=none] (66) at (-5.75, 4.5) {{\tt compile}};
		\node [style=none] (67) at (-7.75, 5) {};
		\node [style=none] (68) at (-3.75, 5) {};
		\node [style=none] (69) at (-3.75, 3.5) {};
		\node [style=none] (70) at (-7.75, 3.5) {};
		\node [style=lalign] (71) at (-7.75, 3.75) {{\tiny\sf pytket Worker}};
		\node [style=none] (73) at (-10.25, 8.75) {};
		\node [style=none] (74) at (-7.25, 8.75) {};
		\node [style=none] (75) at (-7.25, 7.75) {};
		\node [style=none] (76) at (-10.25, 7.75) {};
		\node [style=none] (77) at (-8.75, 8.25) {Circuit};
		\node [style=none] (78) at (-8.75, 7.75) {};
		\node [style=none] (79) at (-6.75, 5) {};
		\node [style=none] (80) at (-5.75, 1) {};
		\node [style=none] (81) at (-5.75, 2) {\tt run};
		\node [style=none] (82) at (-7.75, 2.5) {};
		\node [style=none] (83) at (-3.75, 2.5) {};
		\node [style=none] (84) at (-3.75, 1) {};
		\node [style=none] (85) at (-7.75, 1) {};
		\node [style=lalign] (86) at (-7.75, 1.25) {{\tiny\sf IBMQ Worker}};
		\node [style=none] (87) at (-5.75, 3.5) {};
		\node [style=none] (88) at (-5.75, 2.5) {};
		\node [style=none] (89) at (-5.75, 1) {};
		\node [style=none] (90) at (-5.75, 0) {};
		\node [style=none] (91) at (-7.25, 0) {};
		\node [style=none] (92) at (-4.25, 0) {};
		\node [style=none] (93) at (-4.25, -1) {};
		\node [style=none] (94) at (-7.25, -1) {};
		\node [style=none] (95) at (-5.75, -0.5) {Result};
		\node [style=none] (96) at (-5.75, -1) {};
		\node [style=none] (97) at (-10.25, 8.25) {};
		\node [style=none] (98) at (-10.75, 8.25) {};
		\node [style=none] (99) at (-10.75, -0.5) {};
		\node [style=none] (100) at (-7.25, -0.5) {};
		\node [style=none] (101) at (-4.25, -0.5) {};
		\node [style=none] (102) at (-7.25, 8.25) {};
		\node [style=none] (103) at (0, 8.25) {};
		\node [style=none] (104) at (0, -0.5) {};
		\node [style=none] (106) at (0.75, 5.75) {};
		\node [style=none] (107) at (3.75, 5.75) {};
		\node [style=none] (108) at (3.75, 4.75) {};
		\node [style=none] (109) at (0.75, 4.75) {};
		\node [style=none] (110) at (2.25, 5.25) {Circuit};
		\node [style=none] (112) at (0.75, 5.25) {};
		\node [style=none] (113) at (3.75, 5.25) {};
		\node [style=none] (114) at (4.25, 5.75) {};
		\node [style=none] (115) at (8.25, 5.75) {};
		\node [style=none] (116) at (8.25, 4.75) {};
		\node [style=none] (117) at (4.25, 4.75) {};
		\node [style=none] (118) at (6.25, 5.25) {Parameters};
		\node [style=none] (119) at (4.25, 5.25) {};
		\node [style=none] (120) at (8.25, 5.25) {};
		\node [style=none] (121) at (8.75, 5.75) {};
		\node [style=none] (122) at (12.75, 5.75) {};
		\node [style=none] (123) at (12.75, 4.75) {};
		\node [style=none] (124) at (8.75, 4.75) {};
		\node [style=none] (125) at (10.75, 5.25) {Hamiltonian};
		\node [style=none] (126) at (8.75, 5.25) {};
		\node [style=none] (127) at (12.75, 5.25) {};
		\node [style=none] (128) at (4.25, 3.5) {{\tt substitute params}};
		\node [style=none] (129) at (0.75, 4) {};
		\node [style=none] (130) at (7.75, 4) {};
		\node [style=none] (131) at (7.75, 2.5) {};
		\node [style=none] (132) at (0.75, 2.5) {};
		\node [style=lalign] (133) at (0.75, 2.75) {{\tiny\sf pytket worker}};
		\node [style=none] (134) at (4.25, 2.5) {};
		\node [style=none] (135) at (11, 3.5) {{\tt get Paulis}};
		\node [style=none] (136) at (8.5, 4) {};
		\node [style=none] (137) at (13.25, 4) {};
		\node [style=none] (138) at (13.25, 2.5) {};
		\node [style=none] (139) at (8.5, 2.5) {};
		\node [style=lalign] (140) at (8.5, 2.75) {{\tiny\sf custom worker}};
		\node [style=none] (141) at (11, 2.5) {};
		\node [style=none] (142) at (2.25, 4.75) {};
		\node [style=none] (143) at (2.25, 4) {};
		\node [style=none] (144) at (6.25, 4.75) {};
		\node [style=none] (145) at (6.25, 4) {};
		\node [style=none] (146) at (10.75, 4.75) {};
		\node [style=none] (147) at (10.75, 4) {};
		\node [style=none] (148) at (5.25, 0) {{\tt make circuits}};
		\node [style=none] (149) at (2.75, 0.5) {};
		\node [style=none] (150) at (7.5, 0.5) {};
		\node [style=none] (151) at (7.5, -1) {};
		\node [style=none] (152) at (2.75, -1) {};
		\node [style=lalign] (153) at (2.75, -0.75) {{\tiny\sf custom worker}};
		\node [style=none] (155) at (4.25, 0.5) {};
		\node [style=none] (157) at (6, 0.5) {};
		\node [style=none] (158) at (-1.75, -0.5) {};
		\node [style=none] (159) at (3, -2.75) {};
		\node [style=none] (160) at (6.25, -7.5) {};
		\node [style=none] (161) at (6.25, -6.5) {{\tt compute expectation}};
		\node [style=none] (162) at (2.5, -6) {};
		\node [style=none] (163) at (10, -6) {};
		\node [style=none] (164) at (10, -7.5) {};
		\node [style=none] (165) at (2.5, -7.5) {};
		\node [style=lalign] (166) at (2.5, -7.25) {{\tiny\sf custom worker}};
		\node [style=none] (167) at (5.25, -6) {};
		\node [style=none] (168) at (7, -6) {};
		\node [style=none] (169) at (3.75, -4.25) {};
		\node [style=lalign] (170) at (4.25, 2.25) {{\footnotesize\tt circuit}};
		\node [style=lalign] (171) at (11, 2.25) {{\footnotesize\tt list of Paulis}};
		\node [style=none] (172) at (4.75, -8.5) {};
		\node [style=none] (173) at (7.75, -8.5) {};
		\node [style=none] (174) at (7.75, -9.5) {};
		\node [style=none] (175) at (4.75, -9.5) {};
		\node [style=none] (176) at (6.25, -9) {Energy};
		\node [style=none] (177) at (4.75, -9) {};
		\node [style=none] (178) at (7.75, -9) {};
		\node [style=none] (179) at (6.25, -8.5) {};
	\end{pgfonlayer}
	\begin{pgfonlayer}{edgelayer}
		\draw [style=flow, in=90, out=-90, looseness=1.25] (0.center) to (1.center);
		\draw [style=machine ibm] (6.center)
			 to (5.center)
			 to (4.center)
			 to (3.center)
			 to cycle;
		\draw [style=fugaku] (28.center)
			 to (27.center)
			 to (30.center)
			 to (29.center)
			 to cycle;
		\draw [style=flow, in=90, out=-90, looseness=1.25] (32.center) to (33.center);
		\draw [style=fugaku] (70.center)
			 to (69.center)
			 to (68.center)
			 to (67.center)
			 to cycle;
		\draw [style=border] (73.center) to (74.center);
		\draw [style=border] (74.center) to (75.center);
		\draw [style=border] (75.center) to (76.center);
		\draw [style=border] (73.center) to (76.center);
		\draw [style=flow, in=90, out=-90, looseness=1.25] (78.center) to (79.center);
		\draw [style=machine ibm] (84.center)
			 to (83.center)
			 to (82.center)
			 to (85.center)
			 to cycle;
		\draw [style=flow, in=90, out=-90, looseness=1.25] (87.center) to (88.center);
		\draw [style=flow, in=90, out=-90, looseness=1.25] (89.center) to (90.center);
		\draw [style=border] (91.center) to (92.center);
		\draw [style=border] (92.center) to (93.center);
		\draw [style=border] (93.center) to (94.center);
		\draw [style=border] (91.center) to (94.center);
		\draw [style=box] (97.center) to (98.center);
		\draw [style=box] (98.center) to (99.center);
		\draw [style=box] (99.center) to (100.center);
		\draw [style=box] (101.center) to (104.center);
		\draw [style=box] (104.center) to (103.center);
		\draw [style=box] (103.center) to (102.center);
		\draw [style=border] (106.center) to (107.center);
		\draw [style=border] (107.center) to (108.center);
		\draw [style=border] (108.center) to (109.center);
		\draw [style=border] (106.center) to (109.center);
		\draw [style=border] (114.center) to (115.center);
		\draw [style=border] (115.center) to (116.center);
		\draw [style=border] (116.center) to (117.center);
		\draw [style=border] (114.center) to (117.center);
		\draw [style=border] (121.center) to (122.center);
		\draw [style=border] (122.center) to (123.center);
		\draw [style=border] (123.center) to (124.center);
		\draw [style=border] (121.center) to (124.center);
		\draw [style=fugaku] (130.center)
			 to (129.center)
			 to (132.center)
			 to (131.center)
			 to cycle;
		\draw [style=fugaku] (137.center)
			 to (136.center)
			 to (139.center)
			 to (138.center)
			 to cycle;
		\draw [style=flow] (142.center) to (143.center);
		\draw [style=flow] (144.center) to (145.center);
		\draw [style=flow, in=90, out=-90, looseness=1.25] (146.center) to (147.center);
		\draw [style=fugaku] (150.center)
			 to (149.center)
			 to (152.center)
			 to (151.center)
			 to cycle;
		\draw [style=flow, in=90, out=-90] (134.center) to (155.center);
		\draw [style=flow, in=90, out=-90] (141.center) to (157.center);
		\draw [style=flow, in=90, out=-90, looseness=0.50] (158.center) to (159.center);
		\draw [style=fugaku] (163.center)
			 to (162.center)
			 to (165.center)
			 to (164.center)
			 to cycle;
		\draw [style=flow, in=90, out=-90, looseness=1.25] (169.center) to (167.center);
		\draw [style=flow, in=90, out=-90] (141.center) to (168.center);
		\draw [style=border] (172.center) to (173.center);
		\draw [style=border] (173.center) to (174.center);
		\draw [style=border] (174.center) to (175.center);
		\draw [style=border] (172.center) to (175.center);
		\draw [style=flow] (160.center) to (179.center);
	\end{pgfonlayer}
\end{tikzpicture}
  }
  \caption{Energy Evaluation Workflow.}\label{fig:eeval}
\end{figure}
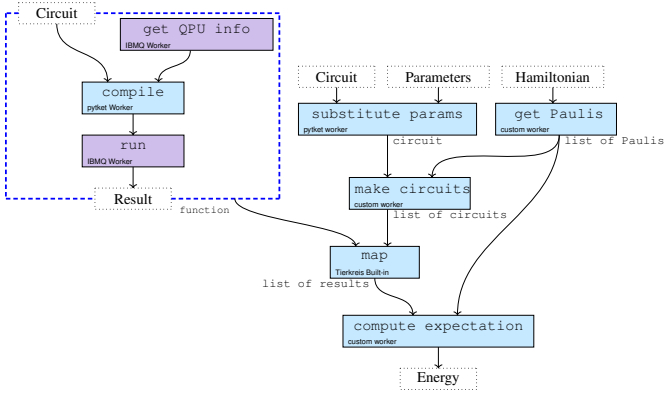

The VQE workflow is considerably more complex; see
Figures~\ref{fig:vqe-wkflow} and \ref{fig:eeval}.  The main element is
the loop construct.  In Tierkreis, \texttt{loop} is a higher-order
operation that takes a workflow as an input and repeats it until a
stopping condition (not shown in this diagram) is
satisfied\footnote{In this experiment we simply stop after 500
  iterations.}.  The loop body may update any (or none) of the input
parameters for the next iteration.  The outputs are emitted to the rest of
the workflow only when the loop terminates.


In our experiment, the loop body implements the SPSA optimiser as
described in Section~\ref{sec:vqe}.  This consists primarily of
computing the gradient at the current point, updating the parameters,
and evaluating the value at the next point.  The updated parameters
are forwarded to the next iteration.  Both the gradient and value
computations perform the same energy evaluation, for which we use
another workflow passed in as an argument.  This inner workflow
generates a list of quantum circuits, and maps another workflow over
this list to compile and run them on ibm\_kobe.  Note that compilation
on ibm\_kobe relies on up-to-date device parameters. While it appears
inefficient to query them in the inner loop, this is needed since
there may be long queue times during the workflow execution, and these
parameters need to be fresh for best compilation.

This example demonstrates the use of many advanced features of
Tierkreis, notably control flow, transparent list handling, and the
ability to dynamically trigger sub-workflows.  The complex
data-handling required is fully automated, which reduces the
possibility of expensive mistakes.

The workflow uses many different kinds of workers.  Alongside the
built-in functions provided by the Tierkreis runtime itself
(e.g. \texttt{map}), we use the \texttt{pytket} worker to manipulate and
compile quantum circuits, and bespoke workers for the Fugaku
installation of Tierkreis, based on \texttt{sqcsub} infrastructure to
access the two QPUs.
The majority of the workers are custom subroutines written for this
experiment.  Such custom workers are written in Python, and included
alongside the workflow definition.  Adding a decorator to the function
definition suffices to declare it as a worker, as shown in
Listing~\ref{lst:tierkreis-task-example}.

The workflow was executed on a login node of Fugaku, which dispatched
the quantum kernels to their appropriate QPU and retrieved the results
for use in the next stage of the workflow.  Little classical
computation was required in this experiment, so no compute nodes of
Fugaku were used.

\subsection{Results}

We run the end-to-end workflow on the $4\times 4$ transverse-field Ising model
with periodic boundary conditions with $h=3J$. For the VQE, we use $D=2$
layers, and $N_S=500$ for each $X/Z$-measurement in $E_\pm$ for a total of 500
iterations. The parameters for QPE are set to $k_\text{max} = 20$, $\lambda =
        0.01\pi/J$ so that all the eigenvalues of $\lambda H$ lie in the interval
$[-\pi, \pi]$. Our implementation uses one ancilla qubit and three scratch qubits for leakage detection.
We use $N_S^{\text{QPE}} = 300$ shots for QPE.

The results of the energy calculation are shown in
Table~\ref{tab:results}.  We simulated 4 different experimental setups,
and obtained hardware results for two of these.  (NB: the pure
VQE result is a byproduct of the workflow design described above.).
Using VQE to ``warm-start'' QPE improves its performance compared to a
classically optimised product state. The hardware results are
consistent with this finding. While VQE+QPE is still significantly
better than VQE alone, the gap on hardware is slightly less than in
simulation. We believe this is due to the compounding error of using a
noisy device to process an imperfectly optimised initial state.

\begin{figure}[thb]
    \centering
    \includegraphics[width=\linewidth]{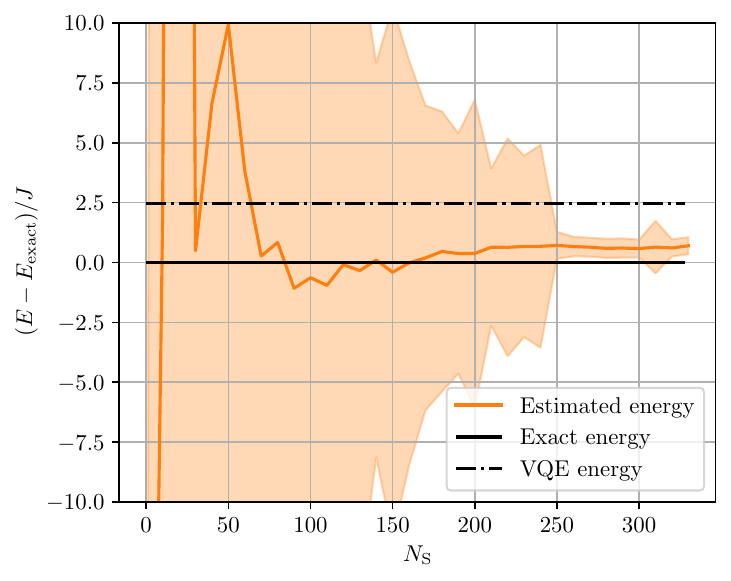}
    \caption{
        The energy estimated by the QPE+VQE experiment on Quantinuum Reimei and ibm\_kobe quantum processors. The opaque orange region represents the statistical uncertainty. For comparison, we also show the energy of the exact ground state ($E_{\rm exact}=-51.45J$) and the VQE state ($E_{\rm VQE}=-48.00J$). The resultant energy estimate is $(E - E_{\rm exact})/J = 0.61\pm0.36$ after $N_S^{\text{QPE}} = 300$ shots.
    }
    \label{fig:energy}
\end{figure}

\begin{table}[bth]
    \centering
    \def\arraystretch{1.5}
    \begin{tabular}{lrr}
        \toprule
        & Hardware & Noiseless \\
        \midrule
        Product & --- & 3.45 \\
        VQE & $2.47\pm0.10$ & $1.25\pm0.15$ \\ \midrule
        Product + QPE & --- & $0.8\pm1.1$ \\
        VQE + QPE & $0.61\pm0.36$ & $0.25\pm0.21$ \\
        \bottomrule
    \end{tabular}
    \caption{\label{tab:results}
        Comparison of energies $(E-E_{\rm exact})/J$ estimated with hardware experiments under various setups.
        For reference, we also show the energies estimated by noiseless simulations.
        ``Product'' stands for the classically optimised product state $(R_Y(\alpha)\ket{0})^{\otimes 16}$, and ``Product+QPE'' stands for the energy estimate by QPE with the product state as the input.
    }
\end{table}

\section{Conclusions} \label{sec:conclusions}

In this work, we have presented two case studies of HPC-quantum hybrid
workflows, using the Fugaku supercomputer and its two QPUs.  
Both of these demonstrations used the Tierkreis workflow management system to
seamlessly integrate the HPC and quantum compute resources.  This
allows us to exploit the power of hybrid compute while managing the
complexity of these mixed workloads.
To enable these workflows, we extended Tierkreis by integrating HPC job submission 
and quantum resource access through dedicated executor mechanisms.
We also developed custom workers to integrate the required software stacks and expose their functionality through typed task interfaces.

In the first application, we show how to generically incorporate
existing software packages into the workflow.  This allows a tighter
software integration (one central control) while maintaining the
flexibility of switching the underlying implementation.  In the second
application, we show the first-of-its-kind end-to-end workflow
including two different quantum modalities.  It shows a scenario where
the advantages of either modality (speed vs. accuracy) can be
exploited to generate meaningful results.  We handle the device
switching through the executor design in combination with the worker
assignment.  
As a result, the workflow itself is agnostic to the underlying hardware,
seamlessly integrating mixed modalities.

Overall, we demonstrate the feasibility and advantages of hybrid
quantum-HPC workflows, challenges of integration, and the potential of
using workflow management systems.  We provide a blueprint for
integrating and developing future hybrid applications; by leveraging a
workflow management system, we separate application development from
the hardware environment and therefore shift the burden of integration
from the scientist to the infrastructure level.

\IEEEtriggeratref{33}
\bibliographystyle{abbrv}
\bibliography{ref}

\end{document}